\documentclass[pdflatex,sn-mathphys-num,referee]{sn-jnl}

\usepackage{graphicx}%
\usepackage{multirow}%
\usepackage{amsmath,amssymb,amsfonts}%
\usepackage{amsthm}%
\usepackage{mathrsfs}%
\usepackage[title]{appendix}%
\usepackage{xcolor}%
\usepackage{textcomp}%
\usepackage{manyfoot}%
\usepackage{booktabs}%
\usepackage{algorithm}%
\usepackage{algorithmicx}%
\usepackage{algpseudocode}%
\usepackage{listings}%
\usepackage{subfigure}
\usepackage{bm}
\usepackage{epsf}
\usepackage{amssymb}
\usepackage{float}
\usepackage{multirow}
\usepackage{epstopdf}
\DeclareGraphicsExtensions{.eps}

\theoremstyle{thmstyleone}%
\theoremstyle{thmstyletwo}%

\theoremstyle{thmstylethree}%

\begin{document}

\title[Computing with symmetry-based qubits]{Deterministic linear-optical computing with symmetry-based qubits}

\author*[1,2]{\fnm{David S.} \sur{Simon}}\email{simond@bu.edu}

\author[2]{\fnm{Christopher R.} \sur{Schwarze}}\email{crs2@bu.edu}

\author[2]{\fnm{Anthony D.} \sur{Manni}}\email{admanni@bu.edu}

\author[2,3]{\fnm{Abdoulaye} \sur{Ndao}}\email{a1ndao@ucsd.edu}

\author[2,4]{\fnm{Alexander V.} \sur{Sergienko}}\email{alexserg@bu.edu}

\affil*[1]{\orgdiv{Dept. of Physics and Astronomy}, \orgname{Stonehill College}, \orgaddress{\street{320 Washington Street}, \city{Easton}, \postcode{02357}, \state{MA}, \country{USA}}}

\affil[2]{\orgdiv{Dept. of Electrical and Computer Engineering \& Photonics Center}, \orgname{Boston University}, \orgaddress{\street{8 Saint Mary's St.}, \city{Boston}, \postcode{02215}, \state{MA}, \country{USA}}}

\affil[3]{\orgdiv{Department of Electrical and Computer Engineering}, \orgname{University of California San Diego}, \orgaddress{\street{9500 Gilman Drive}, \city{La Jolla}, \postcode{92093}, \state{CA}, \country{USA}}}

\affil[4]{\orgdiv{Dept. of Physics}, \orgname{Boston University}, \orgaddress{\street{590 Commonwealth Ave.}, \city{Boston}, \postcode{02215}, \state{MA}, \country{USA}}}

\abstract{A particular type of linear optical multiport, the Grover four-port, has previously been shown to couple the spatial symmetry of a photon to its direction of travel. It is shown here that use of a nonstandard choice of qubit, based on symmetry, allows Grover four-ports to act as compact, low-resource deterministic linear-optical controlled NOT gates, with no post-selection or ancilla measurements required. This approach allows programmable devices, made from Grover multiports in combination with other standard optical components, that can implement multiple different one-, two-, and three-qubit gates, including the Fredkin and Toffoli gates.}


\maketitle


%
%

\section{Introduction}
Although optical platforms for quantum computing have many advantages, such as high speed, low energy consumption, and room temperature operation, the lack of direct photon-photon interactions at normal energies makes controlled gates difficult to implement. One approach is to introduce interactions via nonlinear optical materials \cite{yam,milburn,chuang,dariano,howell2,hutch}, but the resulting interactions between two single photons are very weak at normal levels of nonlinearity. On the other hand, the KLM scheme \cite{klm} provides a means of implementing universal quantum computation with only linear optical elements and detectors. A judicious choice of measurements can create a ``measurement-induced'' nonlinearity.  However, the KLM approach has two drawbacks. First, the resources required tend to grow rapidly with the number of qubits. Second, the operation is probabilistic, working only when particular measurement results have been obtained.  KLM-based and other interference-based approaches have been experimentally implemented in many different ways, for example in Refs. \cite{pitt,ralph,obrien,gasp,hazra}. Various methods have been used to improve scalability, such as by using cluster-state methods \cite{rauss,yor,niel,browne} or avoiding the teleportation used in the original KLM approach \cite{gil}. Up to this point, probabilistic controlled NOT gates implemented experimentally have been able to achieve efficiencies up to about $40\%$ \cite{stoltz} with quantum fidelities of $80$ to $95\%$ \cite{stoltz,lee,chap}.

More recently, proposals have been made to implement CNOT and other controlled gates with Kerr nonlinearities \cite{du} or silicon-vacancy spins \cite{fan} in a manner that promises success probabilities and fidelities approaching 100\%.

A different approach has been to attach multiple qubits to a single photon in such a way that the qubits can interact with each other \cite{cerf,howell,engl}. This allows controlled gates to be implemented in a simpler form, and if extra degrees of freedom are added as a means of additional control, the CNOT gate can be implemented deterministically with linear optics.

Here, we propose an approach that involves encoding qubits into the symmetry or antisymmetry of dual-rail states under horizontal or vertical reflections, independent of the number of particles in the state. We may thus change the value of a symmetry qubit either by applying phase shifts or by adding or removing a photon with appropriate symmetry to the system. In addition, we take advantage of the fact that directionally-unbiased multiports \cite{simon1,osawa2,kim} can introduce correlations between the symmetry and propagation direction of photon states, allowing a coupling of two different qubits.

In some respects, the proposed approach is similar to the use of synthetic dimensions \cite{boada,arg,yuan}, in which spatial dimensions are simulated by alternative variables like orbital angular momentum or energy; this allows, for example, the simulation of gauge fields \cite{celi} or the simulation of electron motions in a spatial lattice by microwave-driven Rydberg energy level transitions \cite{kanungo}. In the present case, qubits are simulated by more abstract quantities that may be attached to a single photon or may be distributed over $n$-photon states of any $n$. Although we focus primarily on the single-photon case, some aspects of the proposed approach work for multi-photon states as well and will be commented on in Section \ref{CNOTsection}.


The approach here, although restricted to qubits, bears similarities to high-dimensional quantum computing, which uses $d$-dimensional qudits in place of $2$-dimensional qubits \cite{fan,he,lanyon,gao,babazadeh,gong,liu}: our approach could be adapted to form a higher-dimensional Hilbert space by, for example, taking the pair of symmetry and direction qubits defined below to form a single four-dimensional qudit. 

In Section \ref{multiport}, we first introduce the qubits and give a brief review of directionally unbiased multiports. We then look in Section \ref{simplegates} at systems in which two qubits of interest are encoded onto the \emph{same} photon, but reflect symmetry about different axes, allowing construction of a broad set of one- and two-qubit single-photon gates. A brief discussion of two-photon symmetry states is given in Section \ref{CNOTsection}. In Section \ref{threequbit}, we return to single-photon states, but now adding couplings between the symmetry states and polarization, allowing the implementation of a range of three-qubit gates, including Fredkin and Toffoli gates. Conclusions are discussed briefly in Section \ref{conclude}.

The main result of this paper is that a broad range of two- and three-qubit operations can be performed in a simple manner, with all qubits encoded on a single photon. Extending this to scalable multiphoton quantum computing is more problematic, as is discussed in section 4, but may be possible with the addition of nonlinear optical components or other extensions of the current formalism.

The approach here is universal for any computation involving no more than three qubits. It might be possible to extend this universality to slightly larger numbers of qubits by incorporating more degrees of freedom, for example angular momentum or time bins. However, to extend to large numbers of qubits in a scalable manner, it is likely that the current approach would need to be combined with other approaches, such as nonlinear materials or KLM methods.

Throughout the following, care must be taken to distinguish, for example, between \emph{two-photon} states and \emph{two-qubit} states, since a single photon may be carrying multiple qubits, and conversely, a single qubit may be distributed over a multiple-photon state.

\section{Photonic qubits and unbiased multiports}\label{multiport}

First, consider a dual rail system.  Label left- and right-moving states on the upper rail as $|a\rangle_{L/R} =\hat a_{L/R}^\dagger |0\rangle$. States on the lower rail are labeled $|b\rangle_{L/R} =\hat b_{L/R}^\dagger |0\rangle$. We will generally be dealing with states that are \emph{distributed over both rails, rather than localized on one}. In particular, define the \emph{symmetrically} and \emph{anti-symmetrically} distributed states,  \begin{eqnarray}|S\rangle_{L/R} &=&{1\over \sqrt{2}}\big( |a\rangle +|b\rangle)_{L/R}\\ &=&  {1\over \sqrt{2}}\big( \hat a_{L/R}^\dagger |0\rangle +\hat b_{L/R}^\dagger|0\rangle)\\ |A\rangle_{L/R} &=& {1\over \sqrt{2}}\big( |a\rangle -|b\rangle)_{L/R}\\ &=& {1\over \sqrt{2}}\big( \hat a_{L/R}^\dagger |0\rangle -\hat b_{L/R}^\dagger|0\rangle).\end{eqnarray}

There are several binary variables associated with each photon that can be used to define qubits, but for the moment we focus on two: (i) The travel direction: for the $j$th photon, define a bit value ${\cal D}_j$, where ${\cal D}_j=0$ for right-moving photons and ${\cal D}_j= 1$ for left-moving. (ii) The symmetry under interchange of the two rails: the symmetry qubit is defined by ${\cal S}_j=0$ for the symmetric state and ${\cal S}_j=1$ for the antisymmetric state.

\begin{figure}
\centering
\includegraphics[totalheight=.5in]{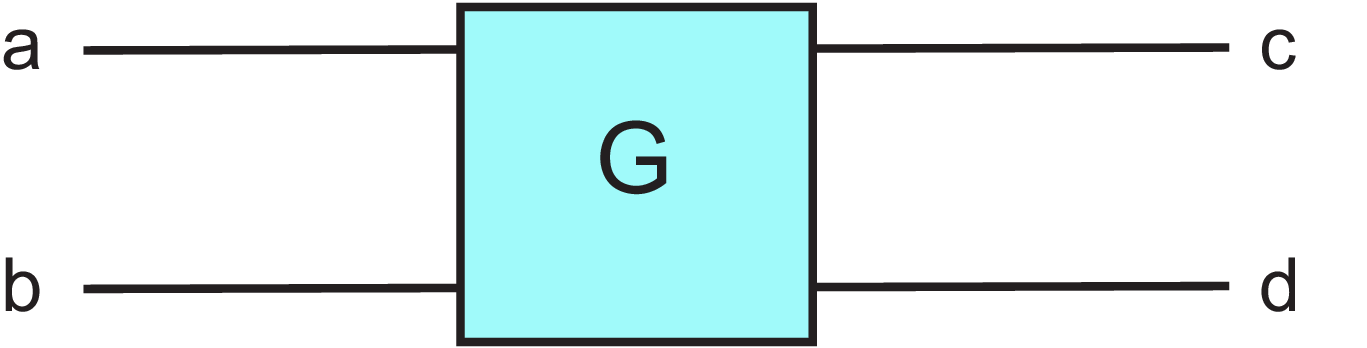}
\caption{A directionally unbiased Grover four-port. All four lines can serve as both input and output. Input at any one port produces equal amplitude output at all four ports, with an extra $\pi$ phase shift at the input port relative to the others }
\label{Ggatefig}
\end{figure}

The reason for emphasizing these two variables is that they can be coupled to each other using a Grover multiport \cite{simon1,osawa2,osawa1,osawax,kim}. (We will see in Section \ref{threequbit} that they can be coupled to polarization, as well.) Directionally-unbiased optical multiports are $n$-port devices that generalize beam splitters in two ways: the number of ports can be \emph{any} integer $n\ge 3$, and the output can exit any of the $n$ ports, including the original input port. The directionally-unbiased three-port has been demonstrated experimentally in bulk \cite{osawa2} and fiber \cite{kim} implementations. The four-port has been implemented in bulk optics experiments \cite{schw2} and via a pair of Y-couplers \cite{ycoupler}. Here, we use the four-port version (Fig. \ref{Ggatefig}), which will be referred to as a Grover multiport due to the fact that it provides a physical implementation of the Grover coin operation used in quantum walks and search algorithms \cite{grover,kempe}. Grover multiports have recently been found useful for distributing entangled states through networks \cite{osawa3,osawa4} and
for expanding the capabilities of interferometers \cite{simonx,schw1,schw2,wei,manni}.

Here, we represent each Grover multiport as a four-port device, in which each input line doubles as an output line. But it can also be mapped into a feed-forward eight-port of the type considered in \cite{zur}.

In a basis labeled by the four ports, the action of the Grover multiport is given by the matrix $U$:
\begin{equation}U={1\over 2}\left( \begin{array}{cccc} -1 & 1 & 1 & 1 \\ 1 & -1 & 1 & 1 \\ 1 & 1 & -1 & 1 \\ 1 & 1 & 1 & -1 \end{array}\right) .\end{equation}

The Grover multiport preserves the symmetry of incoming states, but causes a \emph{change in direction that is conditioned on the symmetry} \cite{cluster}:  \begin{equation}U:|S\rangle_R \to |S\rangle_R ,\qquad |A\rangle_R \to -|A\rangle_L .\end{equation} Here, if a right-moving incoming state is on lines $a$ and $b$ of Fig. \ref{Ggatefig}, then the outgoing state will be on lines $c,d$ for the symmetric case and $a,b$ for the antisymmetric case. (Left-moving input states are affected similarly.) More explicitly,
\begin{eqnarray}{1\over \sqrt{2}} \big( \hat{a}^\dagger +\hat{b}^\dagger\big)_R|0\rangle &\to & {1\over \sqrt{2}} \big( \hat{c}^\dagger +\hat{d}^\dagger\big)_R|0\rangle \\
{1\over \sqrt{2}} \big( \hat{a}^\dagger -\hat{b}^\dagger\big)_R|0\rangle &\to & -{1\over \sqrt{2}} \big( \hat{a}^\dagger -\hat{b}^\dagger\big)_L |0\rangle .\end{eqnarray}
This coupling of symmetry with direction leads to interference that can mimic an interaction between photons, as was demonstrated by the photon-clustering effect discussed in Ref. \cite{cluster}.

For a two-photon state (or more generally, for $n$ photon states), the qubit variables defined above are generalized by adding the qubit values of the individual photons, modulo 2:
${\cal D} =\oplus {\cal D}_j , {\cal S} =\oplus {\cal S}_j$, where $\oplus$ is mod 2 addition and the sum is over all photons. So, for example, the right-moving two-particle states $|S\rangle_R|S\rangle_R$ and $|A\rangle_R|A\rangle_R$ are symmetric overall, ${\cal S}=0$, and have ${\cal D}=0$, while the states $|A\rangle_R|S\rangle_R$ and $|S\rangle_R|A\rangle_R$ are antisymmetric,  ${\cal S}=1$, with ${\cal D}=0$.

${\cal S}$ and ${\cal D}$ are parities under reflection about a pair of orthogonal axes, ${\cal S}$ about the horizontal axis (parallel to the dual rails), and ${\cal D}$ about the vertical axis through the center of the multiport. Although they both measure symmetry or antisymmetry about an axis, we give them different names, ``symmetry'' and ``direction'', in order to clearly distinguish between them. Nevertheless, we will also sometimes refer to both ${\cal S}$ and ${\cal D}$ collectively as ``symmetry qubits'' when the context is clear. The parity qubits of Refs. \cite{klm,gil} are a special case of the symmetry qubits defined here.


\begin{figure}
\centering
\includegraphics[totalheight=1.8in]{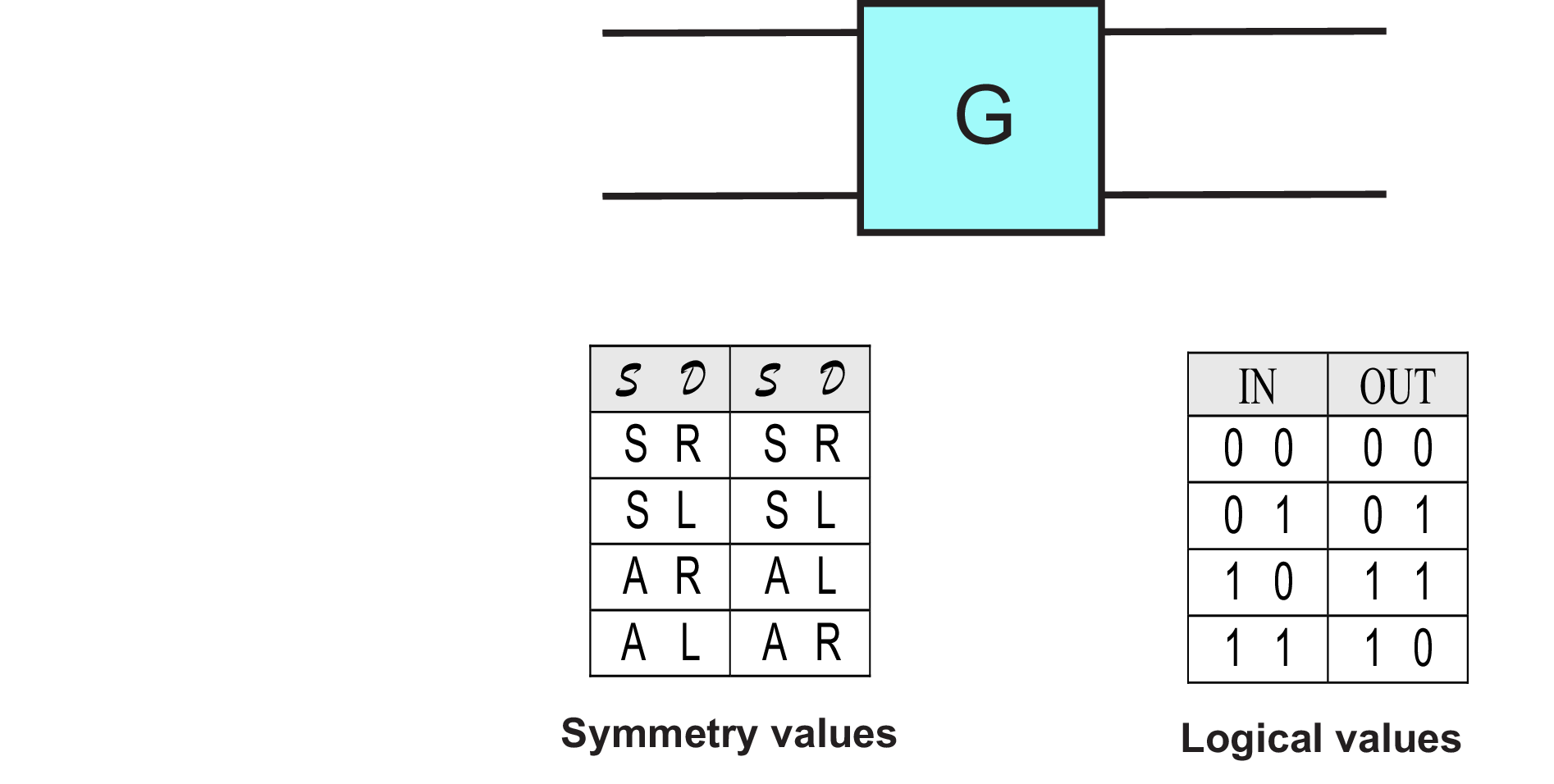}
\caption{The Grover four-port acts as a CNOT on the symmetry and direction qubits. The input and output ${\cal S}$ and ${\cal D}$ qubits in the left-hand table correspond to the logical qubits in the right-hand table. In each table, the first two columns are the input values, the last two columns are output. The CNOT gate acts independently of the number of photons: ${\cal S}$ and ${\cal D}$ can be the symmetry and direction of a single photon, or could be the total symmetry and direction of a multiphoton state}
\label{singlecnotfig}
\end{figure}

\section{Single-photon linear optical symmetry-based gates}\label{simplegates}

In this section we look at one- and two-qubit quantum logic gates that can act on single photons. Here, the symmetry and direction qubits ${\cal S}$ and ${\cal D}$ are assumed to be encoded onto the \emph{same} photon.

\begin{figure}
\begin{center}
\subfigure[]{
\includegraphics[height=1.0in]{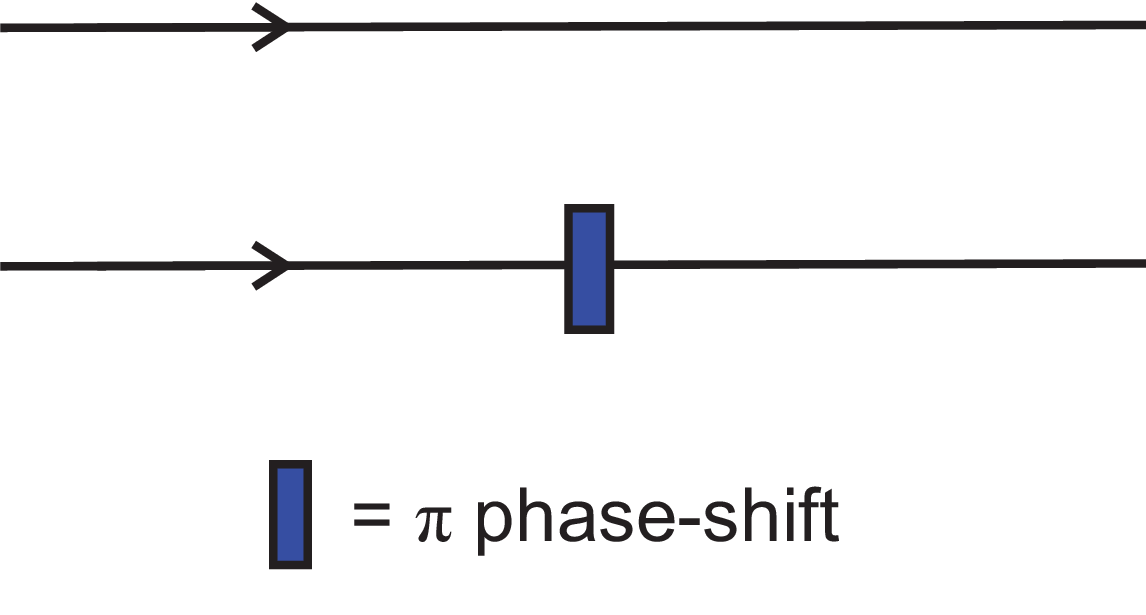}\label{symmflipfig}}
\raisebox{.5in}{\subfigure[]{
\includegraphics[height=.5in]{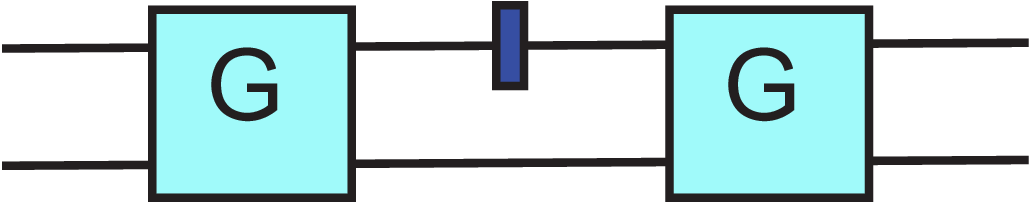}\label{dirflipfig}}}
\caption{NOT gates for the symmetry and direction qubits. (a) Placing a $\pi$ phase shift in one line of a dual rail system will flip the symmetry ${\cal S}$ of the state, $|S\rangle \leftrightarrow |A\rangle$. The phase shift therefore acts as a NOT gate on the single-photon ${\cal S}$-qubit state. (b) A $\pi$ phase shifter between two Grover four-ports similarly acts as NOT gate on ${\cal D}$, while leaving ${\cal S}$ unaffected. A simple two-sided mirror would also achieve the same result}
\end{center}
\end{figure}


As a first example of a single-photon ${\cal S}$-qubit gate, the Grover multiport of Fig. \ref{Ggatefig} acts as a controlled NOT (CNOT) gate, as can be verified by tracing the possible input states through the system, keeping in mind that symmetric states transmit through $G$, while antisymmetric states reflect. (Overall phases of the output states are being ignored here.) Notice that, although we are discussing single-photon states with ${\cal S}$ and ${\cal D}$ attached to the same photon, the CNOT gate in fact also works for multiphoton states, as long as ${\cal S}$ and ${\cal D}$ are taken to be the \emph{total} symmetry and direction of the entire state, as discussed in more detail in Section \ref{CNOTsection}.

\begin{figure}
\begin{center}
\raisebox{.2in}{\subfigure[]{
\includegraphics[height=.8in]{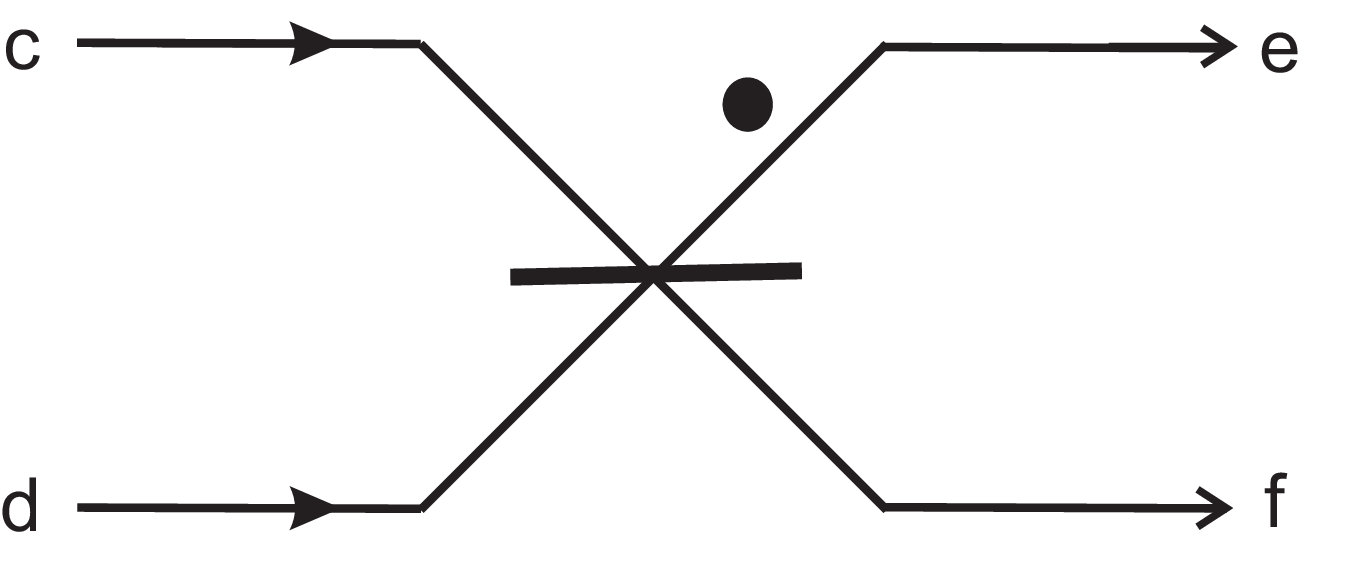}\label{bsfig}}}\quad\quad
\subfigure[]{
\includegraphics[height=1.4in]{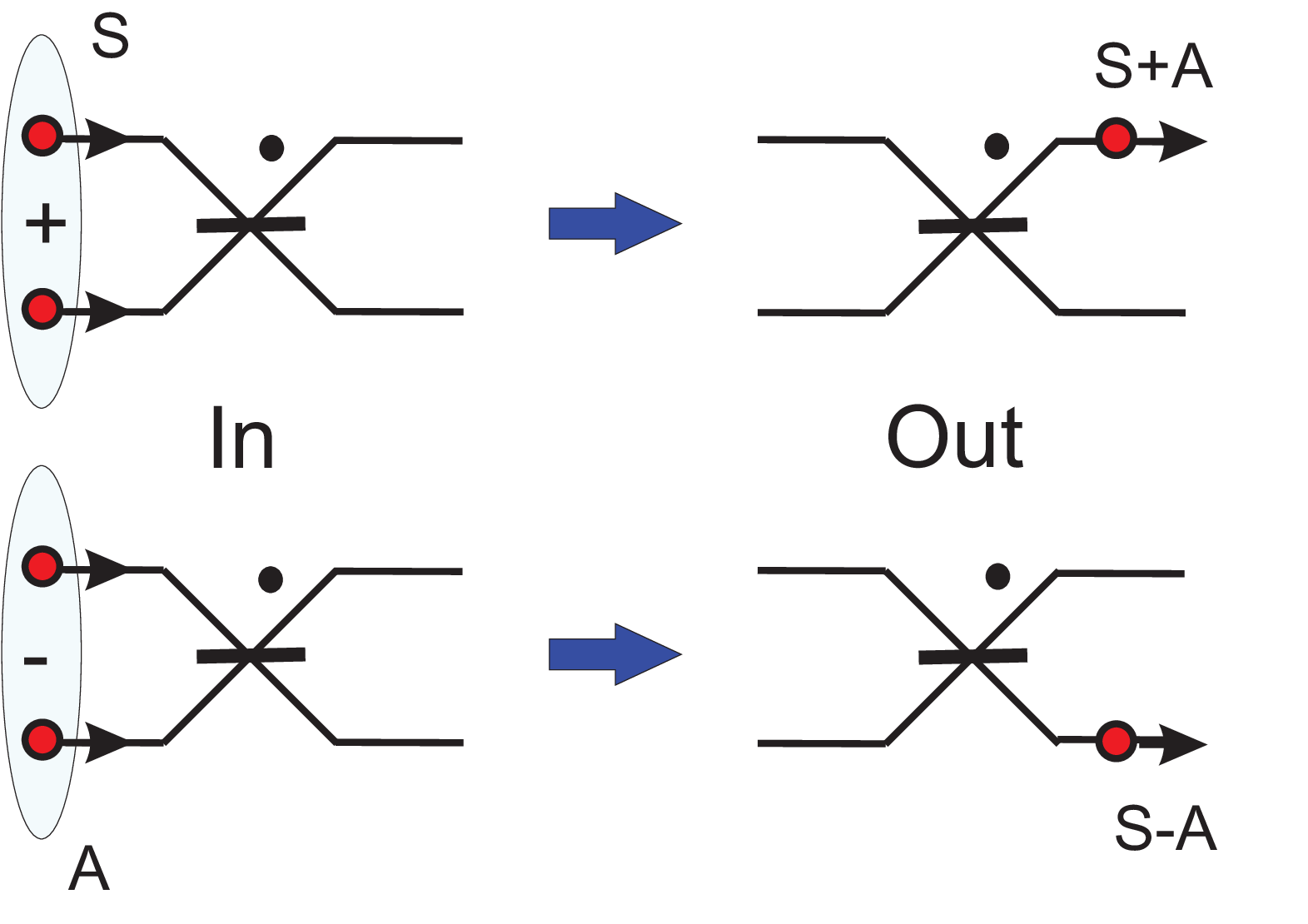}\label{BSeffectfig}}
\caption{A beam splitter (a) can distinguish between symmetric and antisymmetric single-photon states by directing them into different single-rail output modes as shown in (b). The dot denotes the side of the beam splitter at which each single-photon symmetric state exits as a single-rail state. The BS therefore can act as a single-photon Hadamard gate, since the two single-rail modes can be viewed as equal superpositions of symmetric and antisymmetric dual-rail modes, as in (b).}
\end{center}
\end{figure}

Additional symmetry-based gates are easy to envision. For example, a $\pi $ phase shifter placed in one rail of a dual rail system can flip the ${\cal S}$-symmetry of the single-photon state, $ |S\rangle \leftrightarrow |A\rangle$, without altering direction. So, adding such a phase shift to a Grover multiport, as shown in Fig. \ref{symmflipfig}, produces a NOT gate for the symmetry qubit. A NOT gate that acts on direction without affecting ${\cal S}$ is also easy to construct, as shown in Fig. \ref{dirflipfig}.

%

A nonpolarizing beam splitter easily provides a realization of the single-photon ${\cal S}$-qubit Hadamard gate. We assume that the beam splitter of Fig. \ref{bsfig} is a standard dielectric 50/50 beam splitter described by the matrix ${1\over \sqrt{2}}\left( \begin{array}{cc}1 & 1 \\ 1 & -1\end{array}\right) ,$  where the input lines are $c$ and $d$, the outputs are $e$ and $f$.

The action of the beam splitter in Fig. \ref{bsfig} is therefore to turn symmetric or antisymmetric input on lines $c$ and $d$ into single-rail outputs on lines $e$ or $f$ that are equal superpositions of symmetric and antisymmetric states (Fig. \ref{BSeffectfig}),
\begin{eqnarray}BS:|S\rangle_{cd}&\to& |e\rangle \; =\; {1\over \sqrt{2}}\Big(|S\rangle_{ef}+|A\rangle_{ef}\Big)\\ BS: |A\rangle_{cd}&\to& |f\rangle \; =\;  {1\over \sqrt{2}}\Big( |S\rangle_{ef}-|A\rangle_{ef}\Big) .\end{eqnarray} This is exactly the action of a Hadamard gate. The action clearly also works in reverse, allowing single-rail states to be converted into dual-rail symmetry qubits. Notice that there is a sort of complementarity between the Grover four-port and the beam splitter. While the Grover multiport acts on propagation direction and leaves symmetry unchanged, the beam splitter mixes symmetry states while leaving the feed-forward propagation direction unchanged. Combining the action of the two allows control over both degrees of freedom.


\begin{figure}
\begin{center}
\subfigure[]{
\includegraphics[height=1.2in]{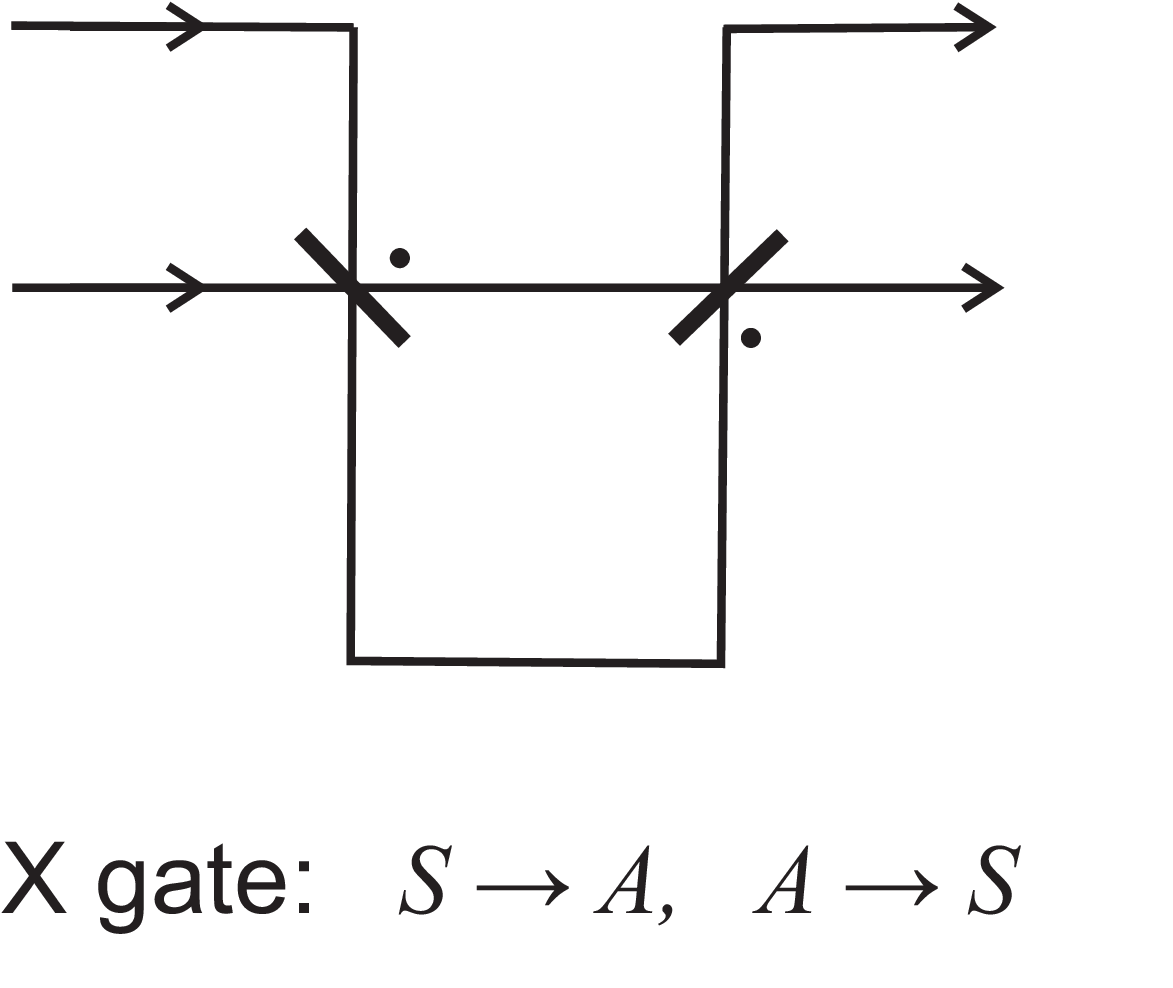}\label{singlexfig}}
\subfigure[]{
\includegraphics[height=1.2in]{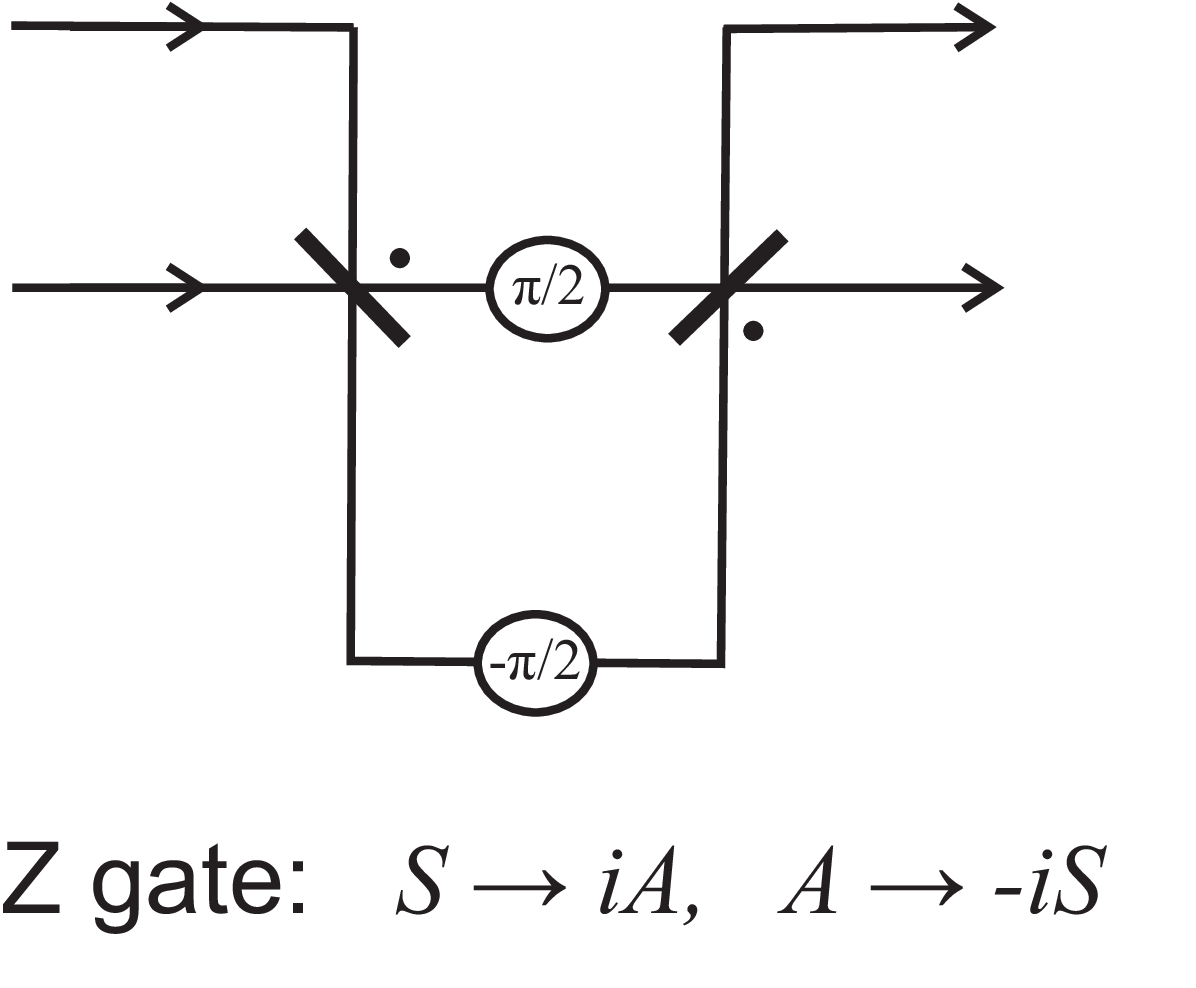}\label{singleyfig}}
\subfigure[]{
\includegraphics[height=1.2in]{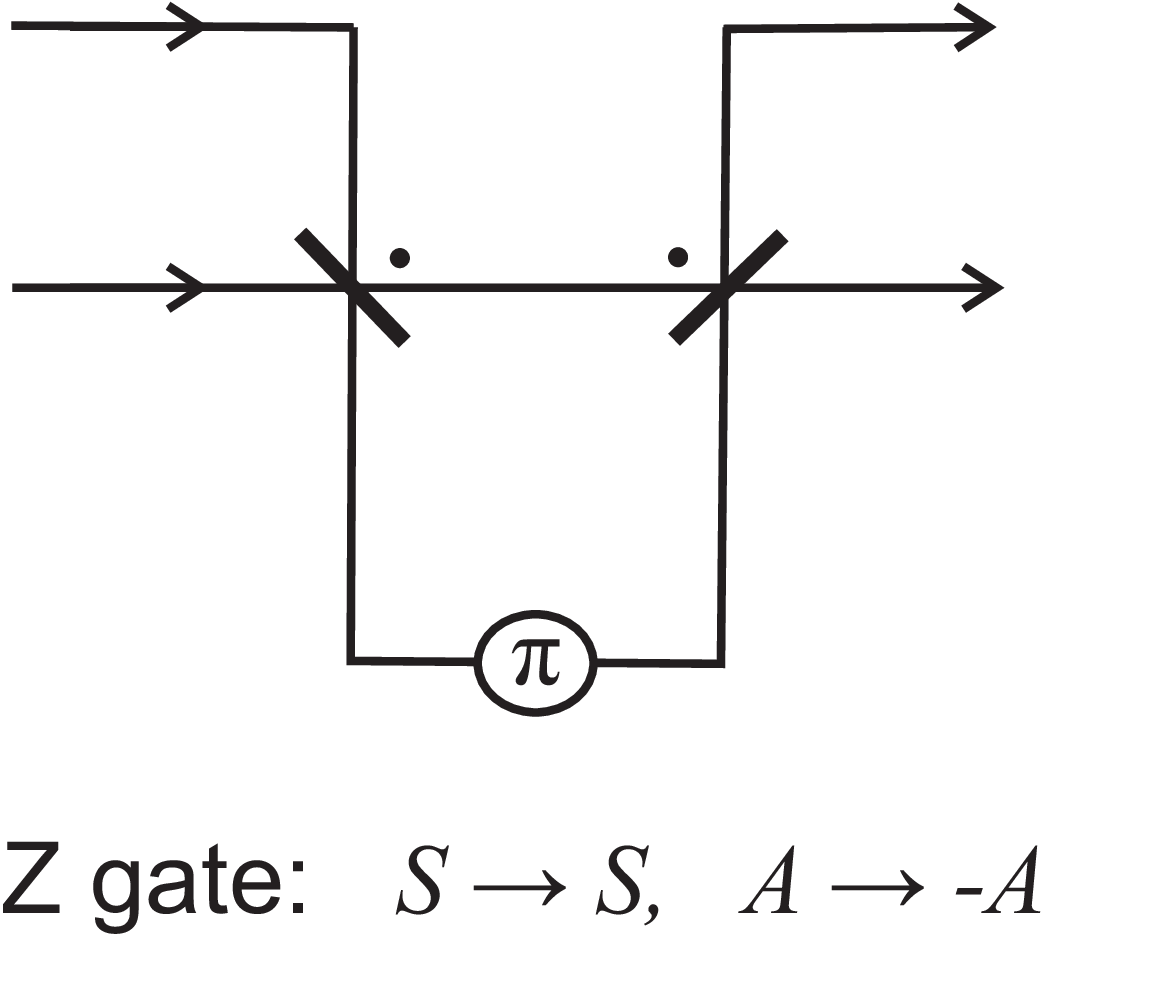}\label{singlezfig}}
\caption{Pauli gates acting on the symmetry qubit, ${\cal S}$. (a), (b), and (c), respectively implement the Pauli $X$, $Y$, $Z$ gates on single-photon states }\label{singlepaulifig}
\end{center}
\end{figure}

Pauli gates acting on ${\cal S}$ can also be implemented, as illustrated in Fig. \ref{singlepaulifig}. In each case, the first beam splitter separates the symmetric and antisymmetric states onto different rails, so that they can be operated on independently, then the second beam splitter converts them back to dual-rail symmetry qubits. Note that the dots on the second beam splitter are in a different position for the $X$ and $Y$ gates relative to the $Z$ gate, indicating an off-diagonal interchange of $S$ and $A$ for the former cases.
\begin{figure}
\begin{center}
\subfigure[]{
\includegraphics[height=0.7in]{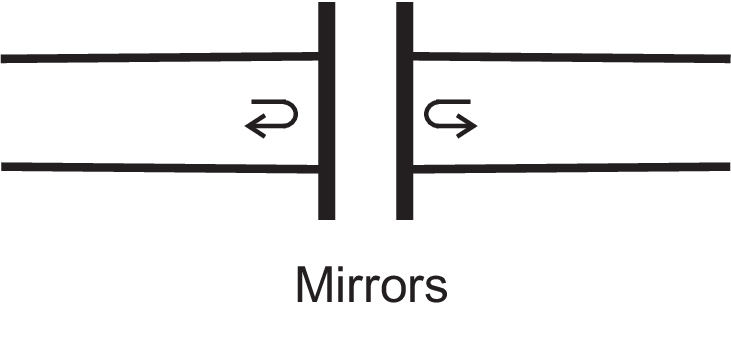}\label{singledxfig}}\qquad
\subfigure[]{
\includegraphics[height=0.7in]{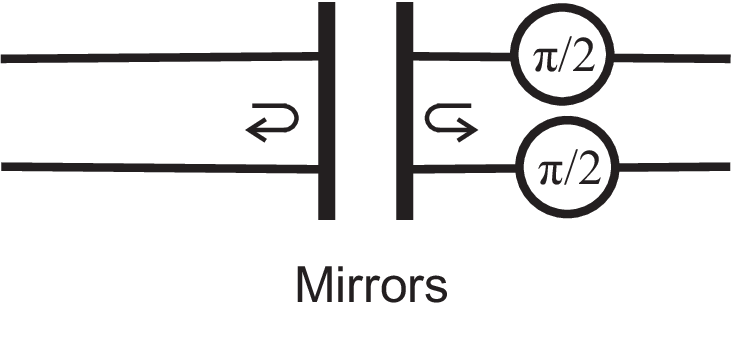}\label{singledyfig}}
\subfigure[]{
\includegraphics[height=1.0in]{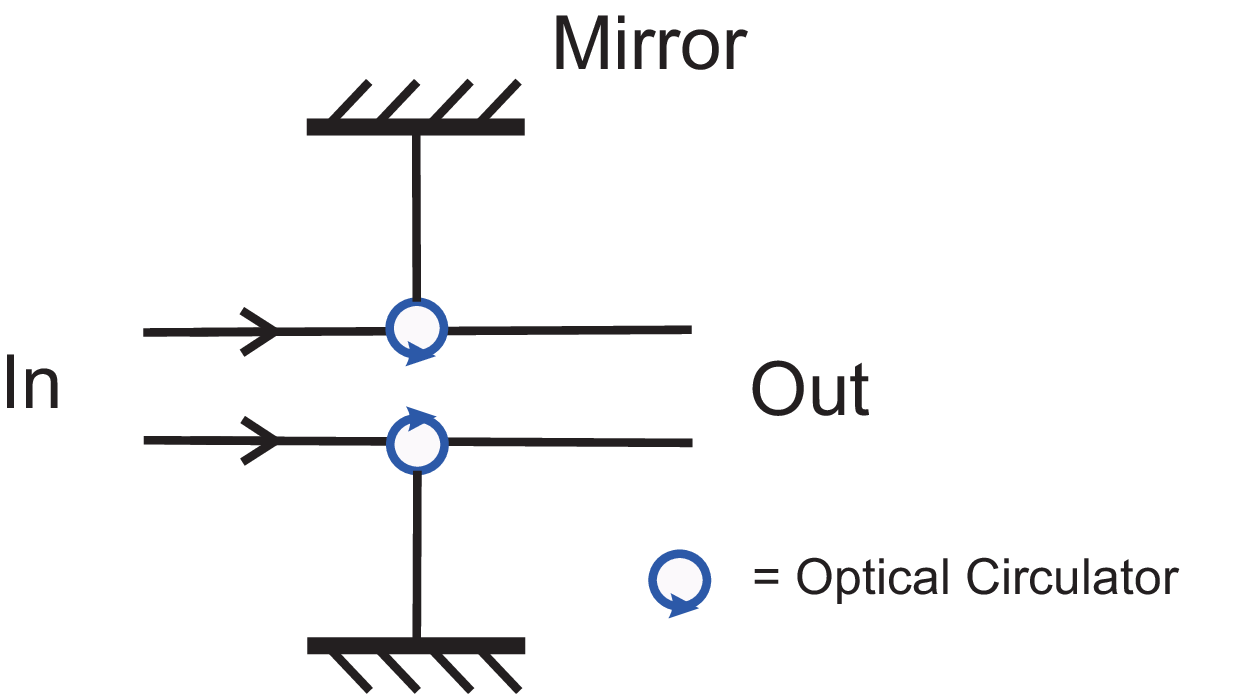}\label{singledzfig}}
\caption{Pauli gates acting on the direction qubit, ${\cal D}$. (a), (b), and (c), respectively implement the Pauli $X$, $Y$, $Z$ gates on single-photon states, up to an overall global phase. For the $X$ gate, each directional qubit is simply reflected by a mirror. For the $Y$ gate, the same is done, but with initially left-moving qubits given an extra $\pi\over 2$ phase shift per pass, relative to the initially right-moving modes.  For the $Z$ gate, the total round-trip phase between circulator and mirror (including any mirror phase) should be a multiple of $2\pi$ }\label{singledpaulifig}
\end{center}
\end{figure}
Similarly, Pauli gates acting on direction are shown in Fig. \ref{singledpaulifig}.

More general phase gates acting on symmetry can also be constructed, by replacing the $\pi$ phase shift in Fig. \ref{singlezfig} by other phase shifts $\phi$, leading to the action $|S\rangle\to |S\rangle $, $|A\rangle \to e^{i\phi} |A\rangle .$

\begin{figure}
\centering
\includegraphics[totalheight=1.2in]{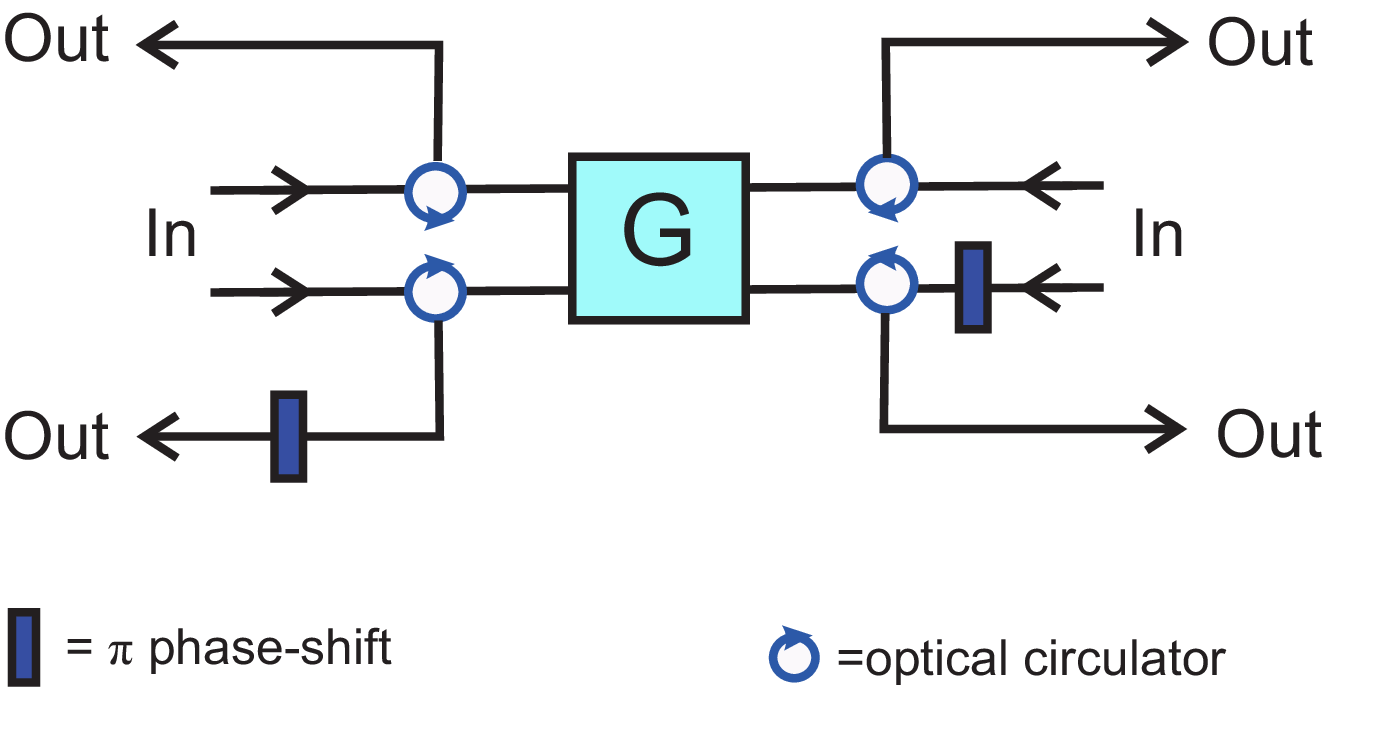}
\caption{A SWAP gate: the output has the values of the symmetry and direction qubits interchanged, ${\cal S}\leftrightarrow{\cal D}$. The optical circulators are used to separate the input from the output states}
\label{swapfig}
\end{figure}

The symmetry and directional information can be inter-converted, ${\cal S}\leftrightarrow {\cal D}$, using an arrangement such as that of Fig. \ref{swapfig}.
This serves as a SWAP gate, interchanging the values of the directional and symmetry qubits.  This arrangement will in fact perform a SWAP operation for ${\cal S}$ and ${\cal D}$ not just on single-photon states, but on $n$-photon states of any $n\ge 1$.

\begin{figure}
\begin{center}
\raisebox{.2in}{\subfigure[]{
\includegraphics[height=.5in]{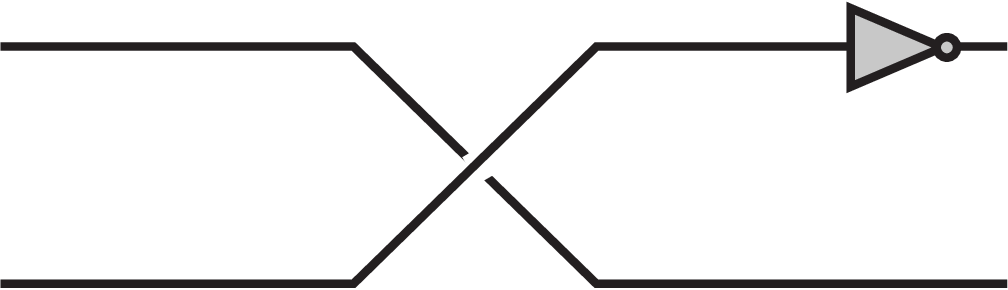}\label{swapnot1fig}}}\quad
\subfigure[]{
\includegraphics[height=.7in]{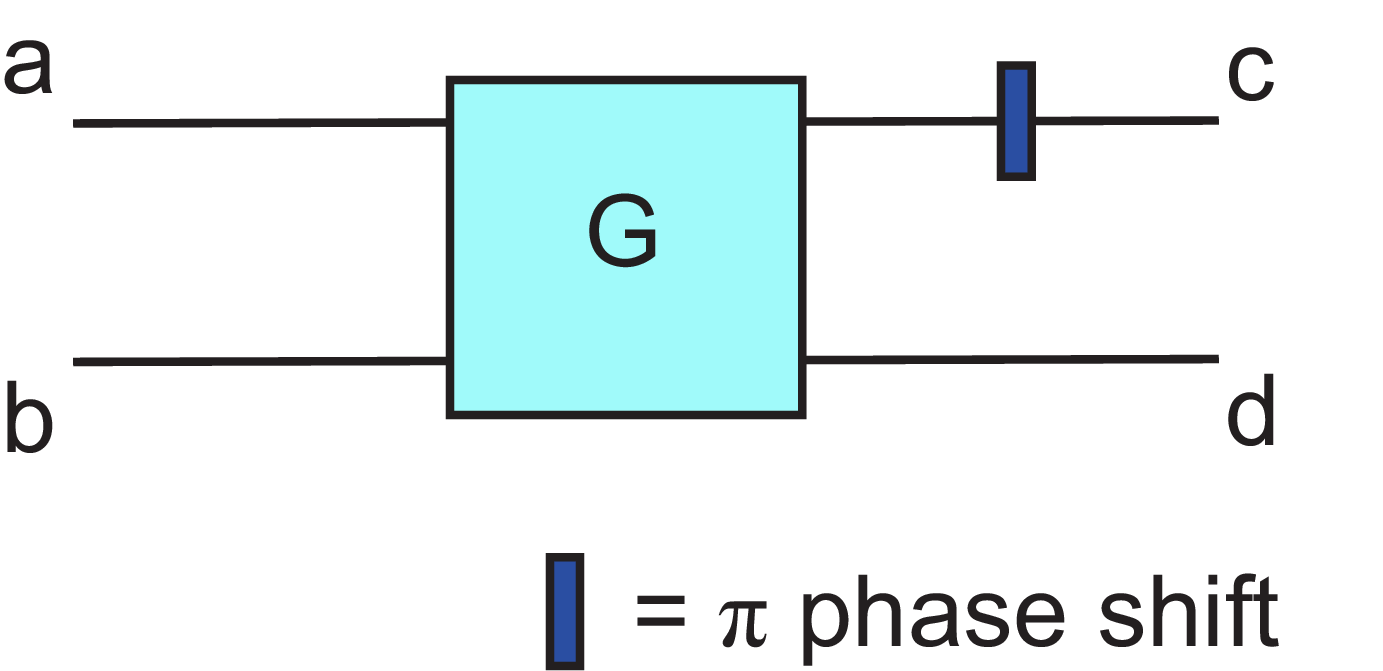}\label{swapnot2fig}}\quad
\caption{(a) Schematic for a circuit that interchanges two qubits and then negates one. (b) A realization of this circuit acting on ${\cal S}$ and ${\cal D}$ qubits }\label{swapnotfig}
\end{center}
\end{figure}

\begin{figure}
\begin{center}
\raisebox{.3in}{\subfigure[]{
\includegraphics[height=.7in]{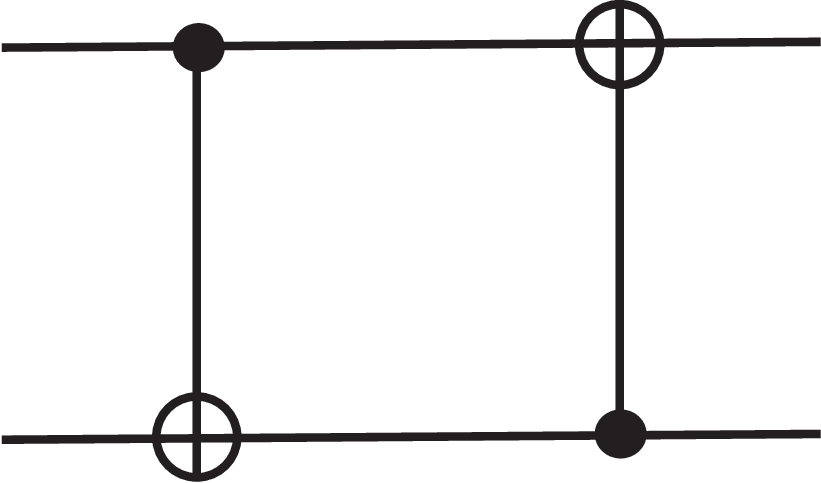}\label{dblcnot1fig}}}\quad\quad
\subfigure[]{
\includegraphics[height=1.1in]{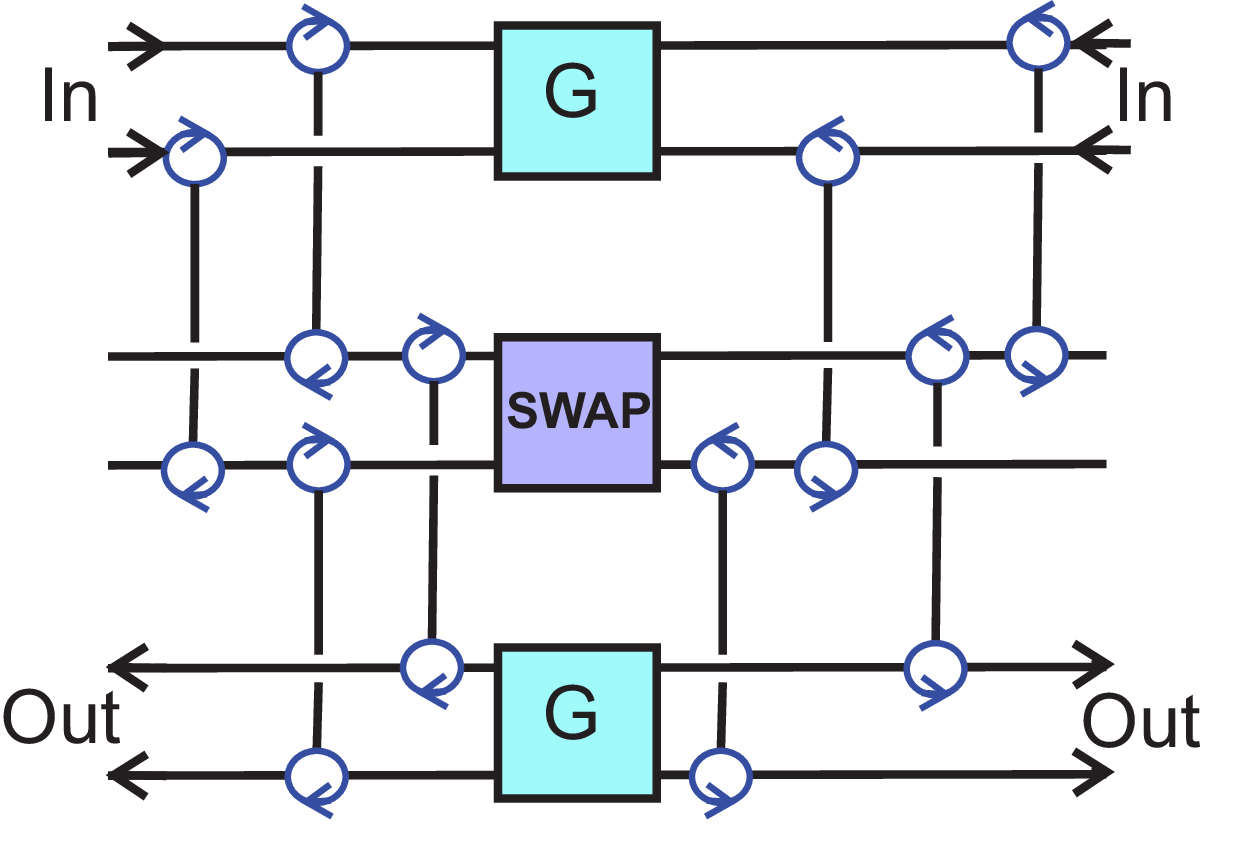}\label{dblcnot2fig}}
\caption{A slightly more complicated circuit example. The circuit shown schematically in (a) can be implemented using ${\cal S}$ and ${\cal D}$ qubits as shown in (b). The blue circles are optical circulators, and the SWAP gate in the middle of the diagram is that of Fig. \ref{swapfig}}\label{dblcnotfig}
\end{center}
\end{figure}

More complicated circuits can be readily constructed that involve these various gates acting in series. A simple example is in Fig. \ref{swapnotfig}, in which a SWAP gate followed by a NOT gate (shown schematically in Fig. \ref{swapnot1fig}) can be implemented as in Fig. \ref{swapnot2fig}.  A more involved example is shown in Fig. \ref{dblcnotfig}.

\section{Two-photon linear-optical gates}\label{CNOTsection}

Using two-photon input states instead of single photons, the qubits defined above might not be attached to each photon individually, but could be distributed over the full two-photon state, i.e. ${\cal S}$ and ${\cal D}$ are the \emph{total} symmetry and direction of all the photons together.  Consider a Grover coin (Fig. \ref{Ggatefig}) acting on these qubits, with two photons entering from left or right, on the dual rails. Each photon is either in a symmetric or antisymmetric state. Imagine for example, one right-moving photon entering from the left in a symmetric state, $|S\rangle_R$. Considering all the possibilities for the second photon, the multiport then acts as follows (ignoring overall phases):
\begin{eqnarray} |S\rangle_R |S\rangle_R  & \to & |S\rangle_R |S\rangle_R  \\
|S\rangle_R |S\rangle_L & \to & |S\rangle_R |S\rangle_L  \\
|S\rangle_R |A\rangle_R & \to & |S\rangle_R |A\rangle_L \\
|S\rangle_R |A\rangle_L & \to & |S\rangle_R |A\rangle_R
\end{eqnarray}
Given the logical bits defined above, this translates into the truth table
\begin{center}
\begin{tabular}{c|c}In&  Out  \\ \hline
$00 $& $0 0$ \\ $01$ & $01 $  \\ $10 $ & $11 $ \\ $11$  & $10 $
\end{tabular}.\end{center}
The first qubit (control) is ${\cal S}$ and the second (target) is  ${\cal D}$.
This is again a standard CNOT truth table.
It is easily checked that all other possible input combinations (for example, replacing the first photon state by $|A\rangle_R$) just reproduce copies of the same table. So we have a completely deterministic two-qubit two-photon controlled gate, with no post-selection or nonlinearity required.

Here, the qubits ${\cal S},\; {\cal D}$ cannot be viewed as attached to either photon individually. In fact, they can be viewed as \emph{differences} between the symmetry or direction of the two photons, since mod-two sums are the same as mod-two differences. So the Grover multiport can also be viewed as a quantum comparator, determining whether properties of different photons are the same or opposite, without actually measuring those properties.

\begin{figure}
\centering
\includegraphics[totalheight=1.4in]{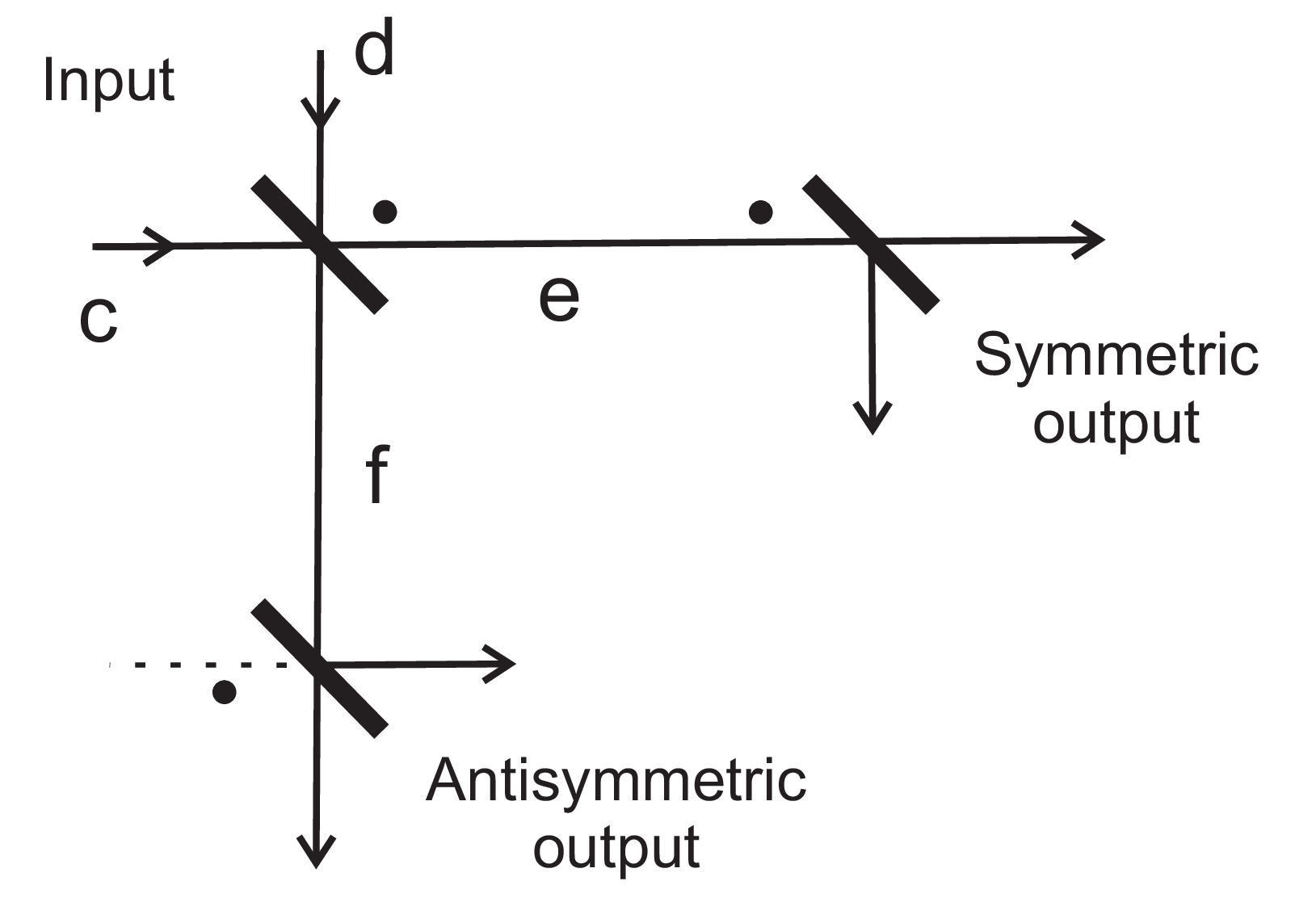}
\caption{Separating symmetric and antisymmetric states into different directions for use in additional gates: single-rail states entering the dotted side exit as symmetric dual-rail states, while those entering at the undotted side exit as antisymmetric states in a different spatial direction }
\label{sepfig}
\end{figure}

Note the important fact that $|S\rangle$ and $|A\rangle$ sharing the same pair of rails can be easily separated from each other for independent processing, simply by passing through a beam splitter (the first BS of Fig. \ref{sepfig}), since the beam splitter acts as follows on two-photon input states:
\begin{equation}|S,S\rangle \to  |ee\rangle , \quad
|S,A\rangle \to |ef\rangle ,\quad
|A,A\rangle \to |ff\rangle .\end{equation}
So states that are overall symmetric (${\cal S}=0$) lead to both photons exiting on the same output line, while overall antisymmetry (${\cal S}=1$) leads to coincidences across both lines. This is essentially just the Hong-Ou-Mandel effect \cite{HOM}.  A set of photon-counting detectors after the beam splitter will easily distinguish the symmetry states by measuring coincidences versus double counts. If instead of detectors, these lines contain additional beam splitters (the two outgoing beam splitters in Fig. \ref{sepfig}), the single-rail states on $e$ and $f$ are converted back into dual rail states that can be directed into different spatial directions for use in additional gates.

%
No nonlinear materials, probabilistic operations, or post-selection are required for the CNOT to operate.
However, the ability to implement deterministic CNOT gates with linear optics does not come for free; the traditional difficulty in implementing the CNOT has simply been moved into a difficulty in implementing other logical components. It is not obvious how to construct a symmetry-qubit Hadamard gate for these multiphoton states.  Nevertheless, the ability to move the difficulty into other components can be a useful means to simplify some types of quantum circuits.



Another difficulty in working with $n$-photon symmetry and direction states for $n>1$ is that although it is easy to project physical states onto symmetry states, the inverse mapping is not unique. For example, the symmetry state with ${\cal S}={\cal D}=0$ could correspond to any of the physical states $|S\rangle_L |S\rangle_L$, $|S\rangle_R |S\rangle_R$, $|A\rangle_L |A\rangle_L$, or $|A\rangle_R |A\rangle_R$ (or to a superposition of them).

Therefore, we avoid these problems by returning in the next section to the case where multiple qubits are encoded on a \emph{single} photon: for single-photon states, invertibility is not a problem and, as shown in Section \ref{simplegates}, the Hadamard gate is easily implemented.

\begin{figure}
\centering
\includegraphics[totalheight=.8in]{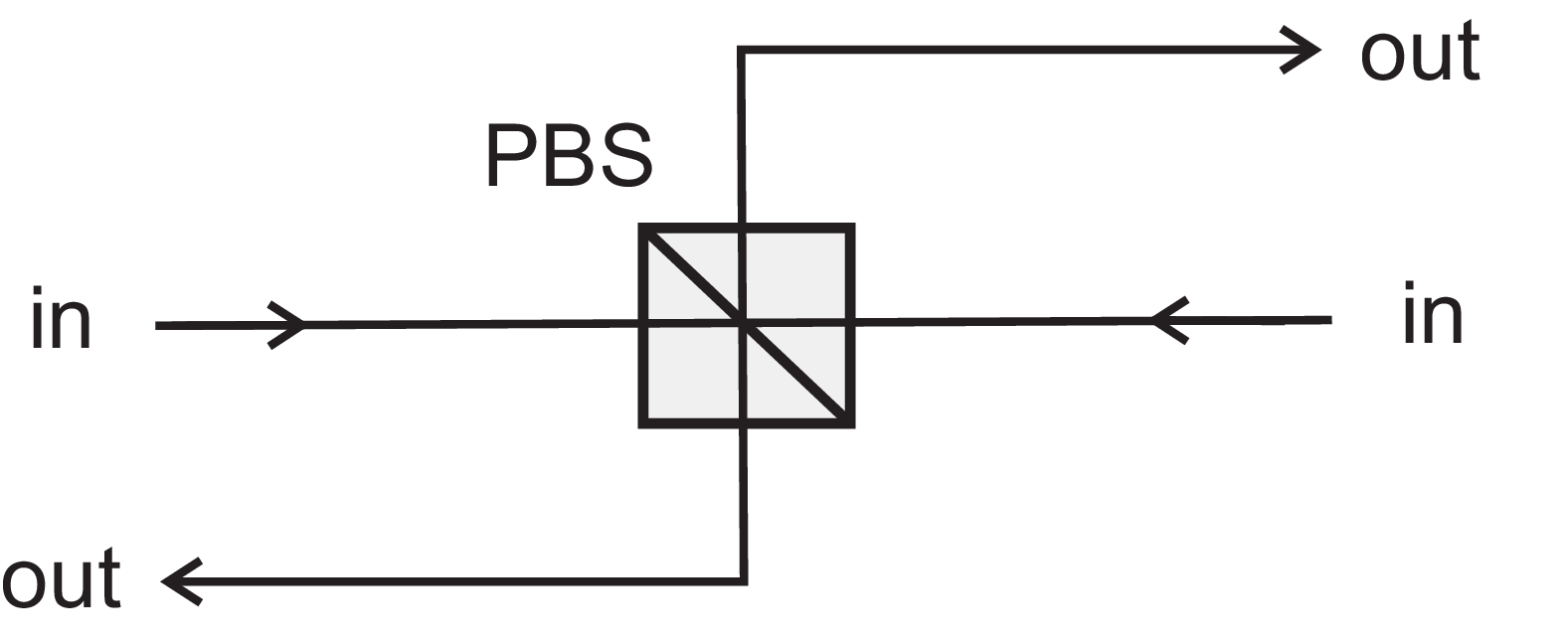}
\caption{A polarizing beam splitter (PBS) can couple polarization information to direction, conditioning whether the output direction flips on the value of the input polarization. Here, it is assumed that the PBS reflects vertically-polarized light and transmits horizontal polarizations}
\label{pbsfig}
\end{figure}

\section{Three-qubit gates}\label{threequbit}

In section \ref{simplegates} we displayed one- and two-qubit gates based on symmetry and direction qubits. Here, we show that by also considering the photon's polarization, a variety of three-qubit gates can be implemented with a single photon. We take the polarization qubit to have value ${\cal P}=0$ for horizontal polarization and ${\cal P}=1$ for vertical.

Recall from Section \ref{multiport} that the Grover multiport couples the symmetry and direction to each other, and can act as a CNOT on direction, with symmetry as control. In a similar manner, a polarizing beam splitter (PBS) can be viewed as coupling direction with polarization. This can be seen in Fig. \ref{pbsfig}, where the PBS acts as a CNOT on direction, with polarization as control: when the photon is horizontally-polarized, the output direction is unchanged, while for vertical polarization the direction flips. (Here, we always assume the PBS reflects $V$ polarization and transmits $H$.) In the figure, the situation is drawn for a single-rail system, but the action is the same on a dual-rail state when a PBS is inserted into both rails.
In a similar manner, a PBS combined with non-polarizing beam splitters can couple polarization to symmetry, using the single-rail to dual-rail mapping discussed in Section \ref{simplegates}.

\begin{figure}
\begin{center}
\subfigure[]{
\includegraphics[height=1.2in]{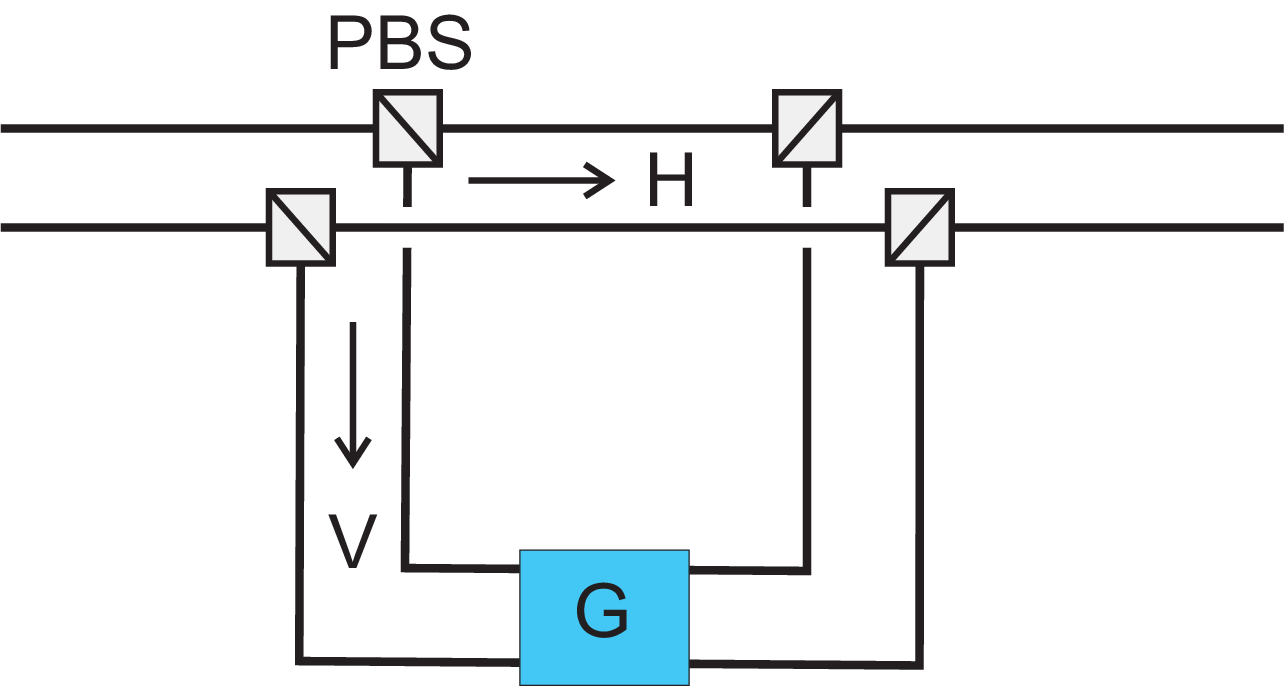}\label{toffig}}\quad
\subfigure[]{
\includegraphics[height=1.5in]{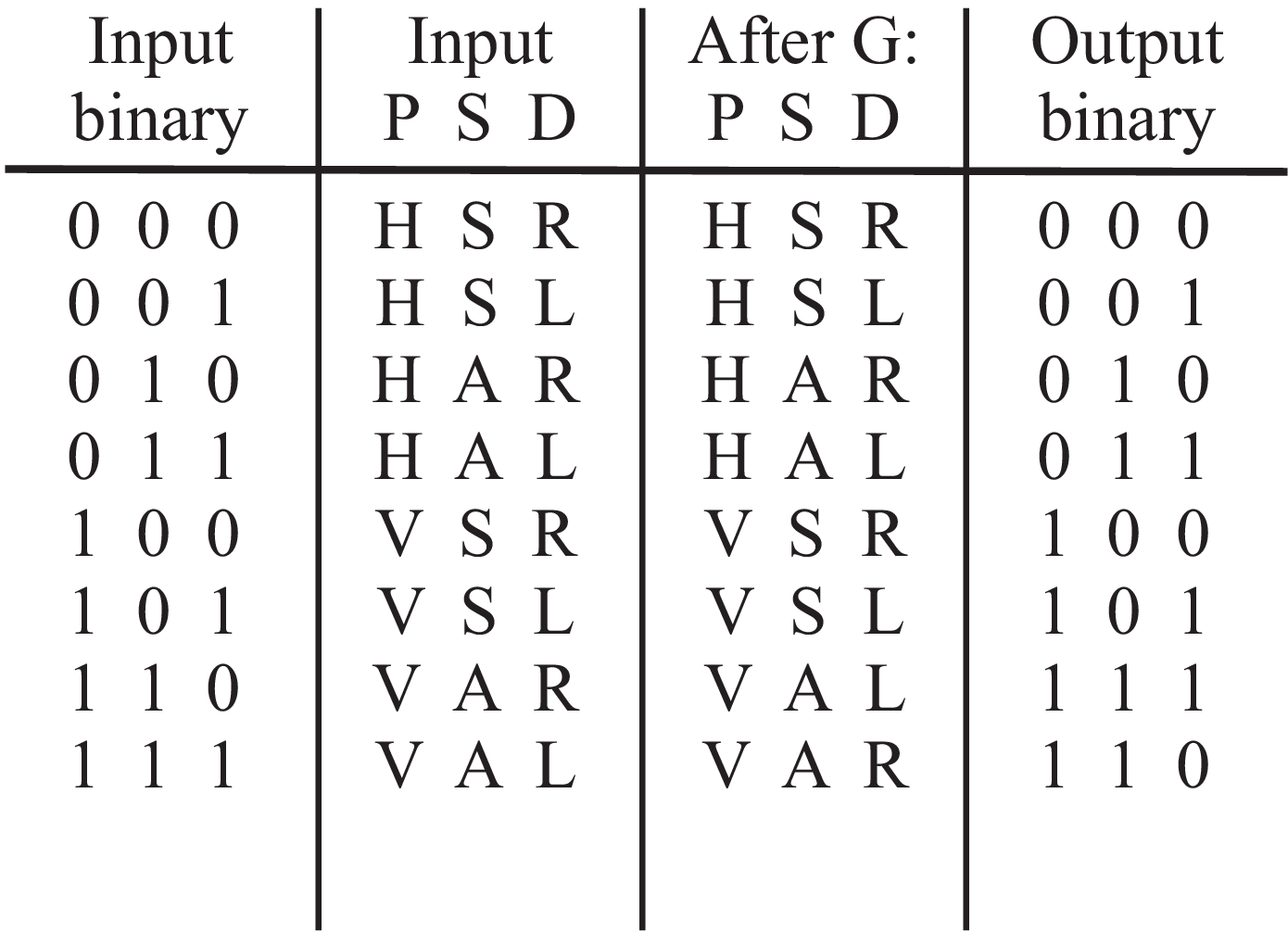}\label{toftabfig}}
\caption{(a) A three-qubit Toffoli gate can be implemented by the circuit shown schematically here, by using symmetry, direction, and polarization qubits. (b) The incoming and outgoing physical and logical qubits are as shown in the table }\label{toffolifig}
\end{center}
\end{figure}

Now we consider three-qubit circuits. As a first example, the circuit of Fig. \ref{toffolifig} implements a three-qubit Toffoli gate. Polarization and symmetry act as the two controls and direction as the target, with all three qubits carried by a single photon.
Note that the Toffoli gate by itself forms a universal gate set, capable of simulating any reversible unitary operation. However, in practice the fact that all of the qubits are attached to a single photon limits the range of computational tasks that can be performed to those involving only a small number of qubits. It is possible that clever use of symmetry qubits attached to multiphoton states may allow effective couplings between qubits on different photons; but the non-invertibility of the mapping between physical qubits and symmetry qubits for two-photon states potentially forms an obstacle to that goal.

\begin{figure}
\begin{center}
\subfigure[]{
\includegraphics[height=1.2in]{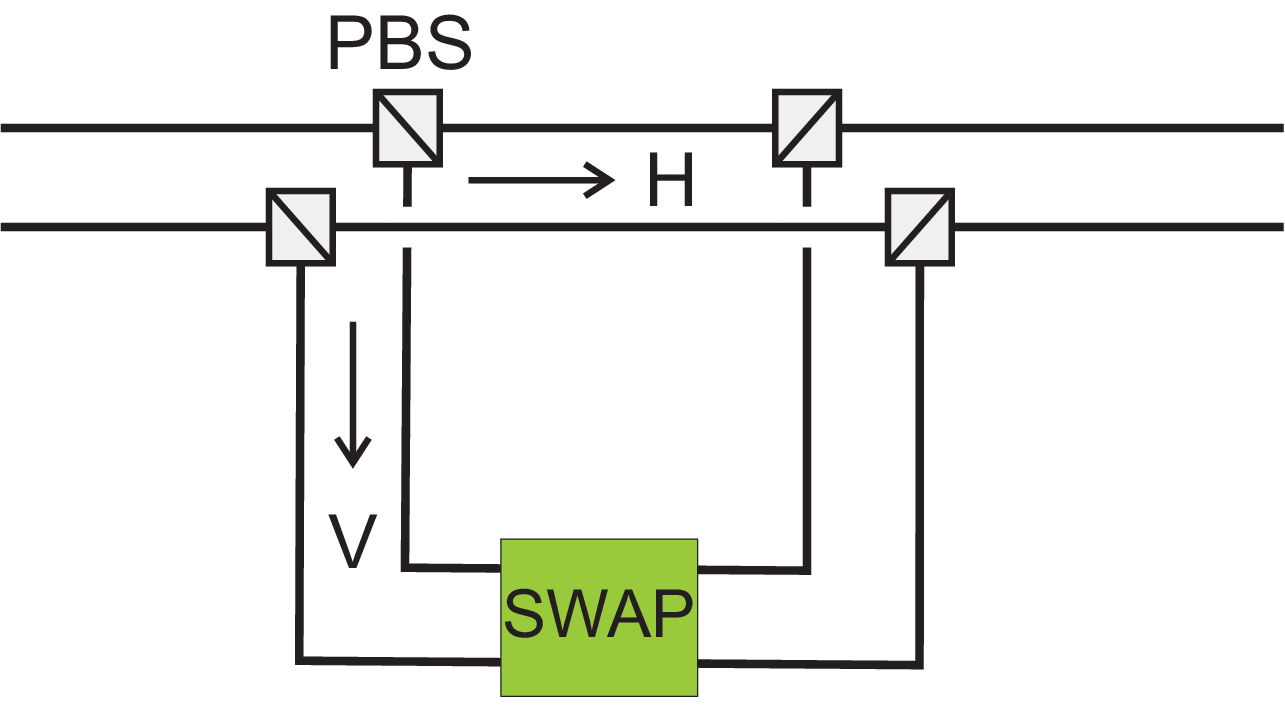}\label{fredfig}}\quad
\subfigure[]{
\includegraphics[height=1.5in]{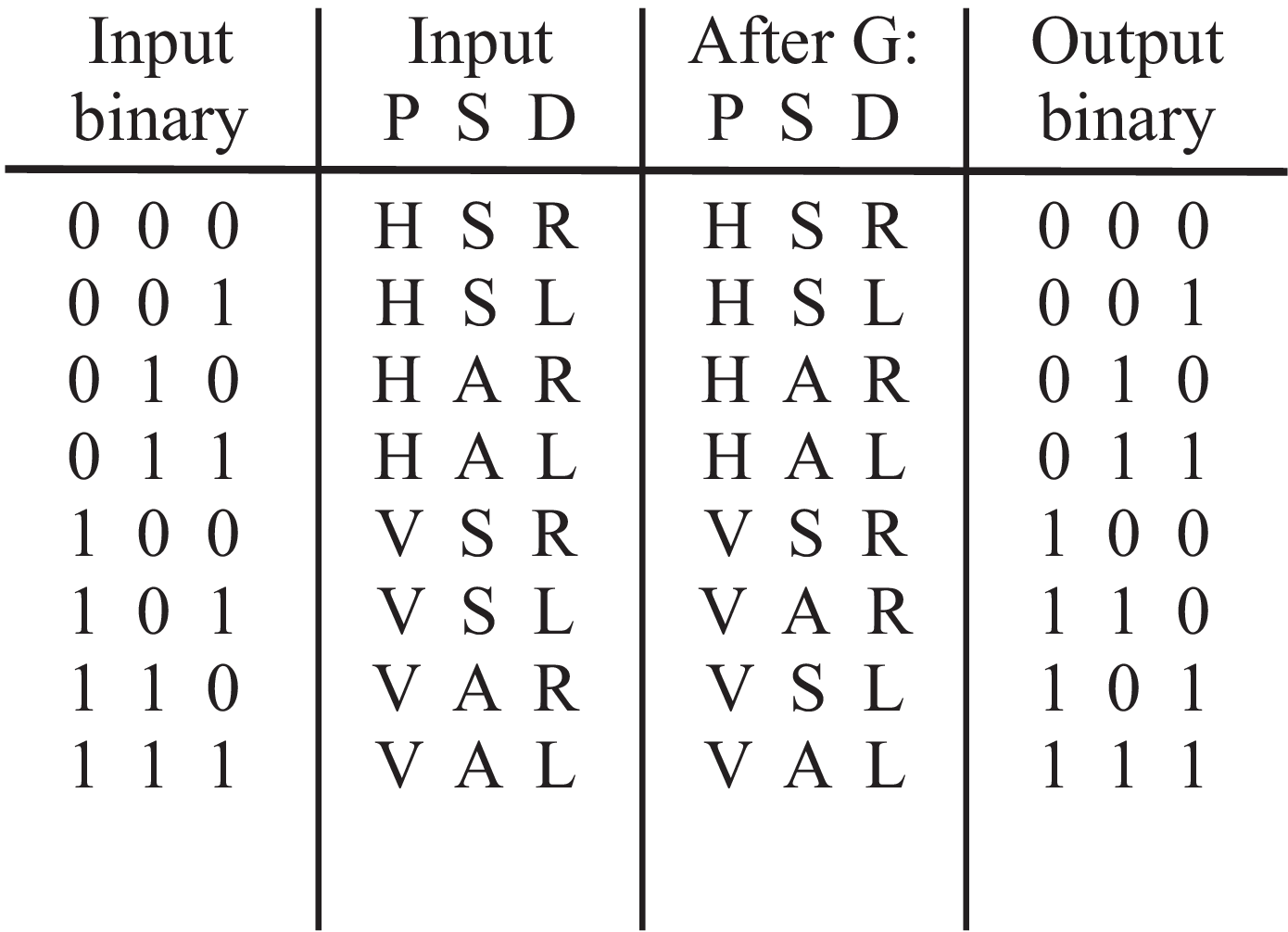}\label{fredtabfig}}
\caption{(a) A three-qubit Fredkin gate can be implemented by the circuit shown schematically here. The SWAP gate is that of Fig. \ref{swapfig}, with optical circulators used direct the SWAP gate input and output into the proper directions. (b) The incoming and outgoing physical and logical qubits are as shown in the table }\label{fredkinfig}
\end{center}
\end{figure}

Replacing the Grover multiport in Fig. \ref{toffig} with the SWAP gate of Fig. \ref{swapfig}, we arrive at the Fredkin, or controlled swap gate, shown in Fig. \ref{fredkinfig}.

\begin{figure}
\centering
\includegraphics[totalheight=1.2in]{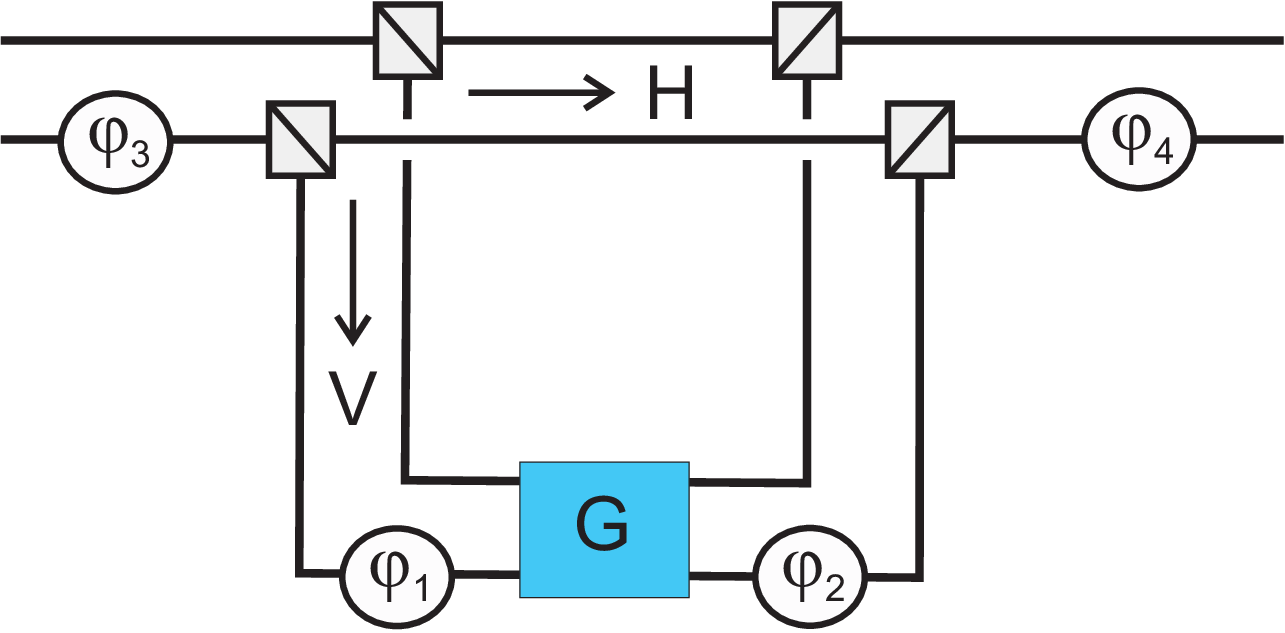}
\caption{A programmable device that can implement multiple different three-qubit logic operations based on whether the phase shifts $\phi_j$ are set to $0$ or $\pi$}
\label{devfig}
\end{figure}

The previous devices can be generalized by adding phase shifts in order to control their operation. For example, consider the device of Fig. \ref{devfig}, which generalizes the Toffoli gate setup. By changing the values of the phases, a variety of different three-qubit tasks can be implemented.  The general case for arbitrary phases is discussed in the appendix. Here we just display a few illustrative examples in cases where the phase shifts $\phi_1$ to $\phi_4$ are either $0$ or $\pi$.

First consider some controlled gates.  When $\phi_1=\phi_2=\phi_3=\phi_4=0$, the device obviously reduces to a Toffoli gate with direction as target: it acts when both controls, symmetry and polarization, equal $1$. But, for example, if $\phi_1=\phi_2=\pi$ and $\phi_3=\phi_4=0$, then the circuit again serves as a Toffoli gate acting on direction, however now acting only when the first control is ${\cal P}=1$ and the second control is ${\cal S}=0$.
Its action is described in the $\left( HSR,HSL,HAR,HAL,VSR,VSL,VAR,VAL \right)$ basis by the matrix \begin{equation}\left( \begin{array}{cccccccc} 1& & & & & & & \\
& 1& & & & & & \\  & & 1 & & & & & \\  & & & 1 & & & & \\  & & & & 0 & -1 & & \\  & & & & -1 & 0 &  & \\  & & & & & & 1  & \\  & & & & & & &  1
\end{array}\right) .\end{equation} (The minus signs, resulting from reflection phases, can be corrected for, in order to make all entries positive, if desired.) As a further example, if $\phi_1=\phi_3=\phi_4=0$ and $\phi_2=\pi$, then there is one control ${\cal P}$ and two targets, with the gate acting only when control value is ${\cal P}=1$. The action on the targets is now more complicated: if ${\cal D}={\cal S}$ the symmetry flips, while if ${\cal D}\ne {\cal S}$ then the direction flips. Effectively, ${\cal S}$ and ${\cal D}$ now each act as the second control for each other.

The device can also act on both $H$ and $V$ polarizations, with different actions on each. For example if $\phi_1=\phi_4=\pi$ and $\phi_2=\phi_3=0$, then vertical states flip their direction if and only if they are symmetric, while all horizontal states reverse their symmetry. This is expressed in matrix form by  \begin{equation} \left( \begin{array}{cccccccc} 0& 0& 1& 0& & & &  \\  0& 0& 0& 1& &  &  &  \\ 1& 0& 0 &0  &  &  &  &  \\
0 & 1 & 0 &0  & & &  &  \\ & &  & & 0 &-1  &  &  \\ & &  & & -1 &0 &  &  \\ & &  & &  &  & 1 &0  \\ & &  & &  &  & 0 &  1
\end{array}\right) .\end{equation}  In contrast, if $\phi_1=\phi_3=\pi$ and $\phi_2=\phi_4=0$, then horizontal states again flip their symmetry, but vertical states flip their direction if and only if they are antisymmetric. The corresponding matrix is \begin{equation} \left( \begin{array}{cccccccc} 0& 0& 1& 0& & & &  \\  0& 0& 0& 1& &  &  &  \\ 1& 0& 0 &0  &  &  &  &  \\
0 & 1 & 0 &0 & &  &  &  \\ & &  & & 1 &0  &  &  \\ & &  & & 0 &1  &  &  \\ & &  & &  &  & 0 & -1 \\ & &  & &  &  &  -1&  0
\end{array}\right) .\end{equation}

The blocks in the matrices for this device can easily be seen to have several symmetries under interchanges of the phases: for example, interchanging $\phi_2\leftrightarrow \phi_4$ and  $\phi_1\leftrightarrow \phi_3$ leaves the lower right block (the action on vertical states) invariant, while the interchanges $\phi_1\leftrightarrow \phi_2$ and  $\phi_3\leftrightarrow \phi_4$ leave the upper left block (horizontal) unchanged.

By allowing the phase shifts to be externally controlled, the device of Fig. \ref{devfig} can therefore be programmed to carry out multiple different three-qubit processes. By changing phases, a variety of different couplings between polarization, symmetry, and direction can be arranged. This can be further generalized by placing additional Grover coins, swap gates, or other devices in the upper and lower branches of the circuit, allowing expansion of the operations that can be performed. In addition, polarization rotators can be added to go beyond the $4\times 4$ block diagonal forms of the examples above. These additional expansions will be explored elsewhere.

\section{Conclusion}\label{conclude}
It has been shown that by defining nonstandard synthetic symmetry-based qubits, it is possible to construct a universal set of deterministic one-, two-, and three-qubit gates with linear optical devices.

Although not examined in detail here, it is easy to envision adding in additional degrees of freedom such as time bins and orbital angular momentum to implement $n$-qubit gates with $n>3$. The main point is that different degrees of freedom, such as symmetry and polarization, can be mixed and matched in order increase the capabilities of a system. This is in contrast to earlier approaches, such as \cite{cerf}, that encoded multiple qubits onto a single photon via a \emph{single} degree of freedom that took on a larger number of values (for example, which-path information in a system with many paths).

Assuming $N$ degrees of freedom are encoded into a single photon, the resources required to carry out $N$-qubit gates grows relatively slowly with $N$. Each added degree of freedom seems to require no more than two additional pairs of beam splitters (one pair on each side of the Grover multiports) to spatially separate that variable from the others, and each of those beam splitters can require at most one new multiport to feed into. So we would expect the resources required for $N$ qubits to scale linearly in $N$. This is borne out by the examples given in earlier sections. The low resources required in this approach can be seen starkly by comparing to analogous gates in other approaches; for example, compare the parsimony of the Toffoli gate of the previous section to the Toffoli gate described in Fig. 3 of \cite{du}, which requires 10 beam splitters and two $X$ gates, plus six swap gates (each with additional beam splitters and $X$ gates embedded inside each swap gate), as well as a cross-Kerr nonlinearity. 

A central message here is that computing capabilities in a specific platform are dependent on the choice of qubit variable, and that by defining more abstract synthetic qubits it may be possible to perform computations with a given platform that cannot be conducted with standard qubit definitions. For example, linear-optical CNOT gates, which are difficult to achieve using polarization or other standard qubit platforms, are easy to implement using symmetry qubits. In the quest for practical quantum information processing capabilities, the search for advantageous synthetic qubits should therefore be pursued in parallel to work on improvement to hardware capabilities.

\bmhead{Acknowledgements}

This research was supported by the Air Force Office of Scientific Research MURI award number FA9550-22-1-0312.

\section*{Declarations}

\begin{itemize}
\item Conflict of interest: The authors declare no conflict of interest.
\item Data availability: No datasets were generated or analyzed during the current study.
\item Author contribution: D.S.S. originated the central concept and wrote the main manuscript text. A.D.M., C.R.S, A.N., A.V.S. contributed additional ideas and gave feedback on methods and interpretation of results. All authors reviewed the manuscript.
\end{itemize}

\begin{appendices}

\section{General form of the three-qubit device}\label{secA1}

\begin{figure}[t!]
\centering
\includegraphics[totalheight=.6in]{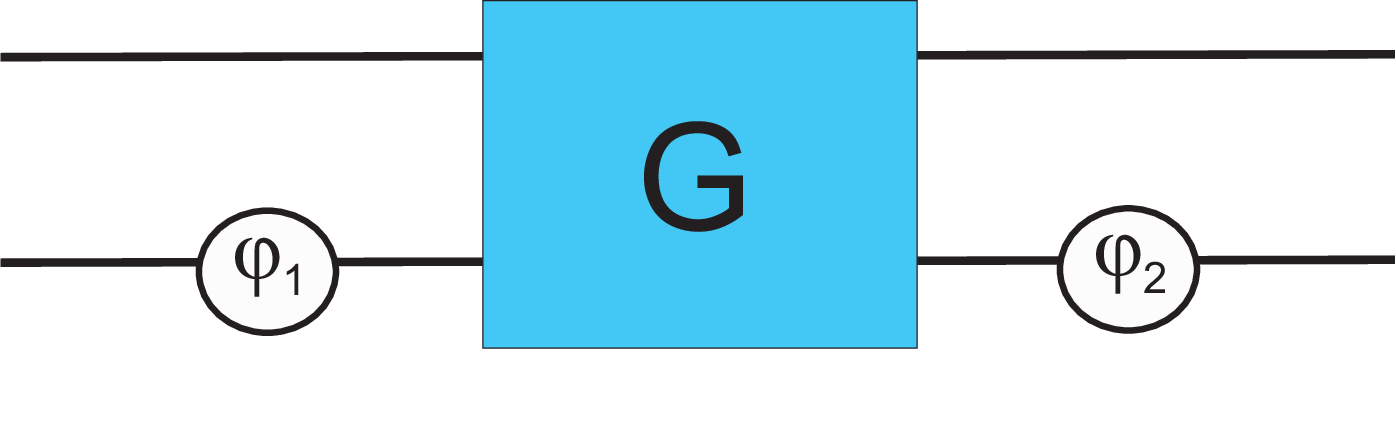}
\caption{The segment of the device in Fig. \ref{devfig} that acts on vertically-polarized states.}
\label{UGUfig}
\end{figure}

Here we briefly outline the general form of the matrix for the circuit for  Fig. \ref{devfig}. Consider the segment of the circuit shown in Fig. \ref{UGUfig}. Working in the basis $\left(SR,SL,AR,AL\right)$, the action of the Grover multiport is \begin{equation}G= \left(\begin{array}{cccc} 1 & & & \\   & 1& & \\ & & 0 &-1 \\ & & -1& 0 \end{array}\right), \end{equation} and each of the phase shifts takes the form \begin{equation}U_j(\phi_j)= e^{i\phi_j/2}\left(\begin{array}{cccc} c_j & 0& -is_j & 0 \\   0& c_j& 0& -is_j \\ -is_j& 0& s_j &0 \\ 0& -is_j& 0& c_j \end{array}\right) , \end{equation} for $j=1,2$.  Here, we have abbreviated $c_j=\cos\phi_j$ and $s_j=\sin\phi_j$.
So the complete segment in Fig. \ref{UGUfig} is given by
\begin{equation} U_2GU_1= e^{i(\phi_1+\phi_2)/2}  \left( \begin{array}{cccc}  c_1c_2 & s_1s_2 & -ic_2s_1 & is_2c_1\\
s_1s_2 & c_1c_2 & is_2c_1 & -ic_2s_1 \\  -is_2c_1 & ic_2s_1 & -s_1s_2 & -c_2c_1\\ ic_2s_1 & -is_2c_1 & -c_1c_2 & -s_1s_2  \end{array}\right)  .   \end{equation}

Tracing each of the possible input states through the system then allows the matrix determining the action of the full system to be constructed. The action will be different on horizontally- and vertically-polarized states (without mixing the polarizations), so the matrix takes block diagonal form:
\begin{equation} U_{total}=\left(\begin{array}{cc}   e^{i\phi^{\prime\prime}}A & 0\\ 0 & e^{i\phi} B    \end{array}\right) ,  \end{equation}
where $\phi={{\phi_1+\phi_2}\over 2}$ and $\phi^\prime={{\phi_3+\phi_4}\over 2}$, and where the blocks $A$ and $B$, representing the action on horizontal and vertical states respectively, are  given by
\begin{eqnarray} A&=&\left(\begin{array}{cccc}  \cos\phi^{\prime\prime} & 0 & -i\sin\phi^{\prime\prime} & 0\\
0& \cos\phi^{\prime\prime} & 0 & -i\sin\phi^{\prime\prime}\\   -i\sin\phi^{\prime\prime} & 0 & \cos\phi^{\prime\prime} & 0 \\  0 & -i\sin\phi{\prime\prime} & 0 & \cos\phi^{\prime\prime}   \end{array}\right)  , \\
B &=&\left(\begin{array}{cccc}  cc^\prime & s^{\prime\; 2} e^{-i\Delta \phi} & -isc^\prime & ic^\prime s^\prime e^{-i\Delta \phi}\\ s^2 e^{i\Delta \phi} & c c^\prime & isc\; e^{i\Delta \phi} & -ics^\prime \\
-ics^\prime & is^\prime c^\prime e^{-i\Delta \phi} & -ss^\prime  & -c^{\prime \; 2}e^{-i\Delta \phi} \\ isc\; e^{i\Delta \phi} & -isc^\prime & -c^2 e^{i\Delta \phi} & -ss^\prime
 \end{array}\right) .\end{eqnarray} Here we have defined $c=\cos\phi$, $c^\prime =\cos \phi^\prime$ (and similar for the sines), as well as $\Delta \phi =\phi-\phi^\prime$ and $\phi^{\prime\prime}={{\phi_3+\phi_4}\over 2}$.

\end{appendices}

\bibliography{symmbibliography}
\end{document}